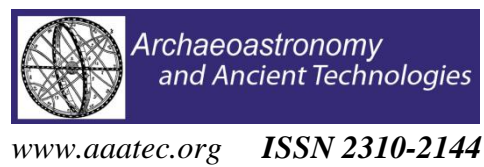

# Astronomical Interpretation of the Signs on the Vessel of the Bronze Age (Central Donbass)

**Larisa N. Vodolazhskaya[1], Anatoliy N. Usachuk[2], Mikhail Yu. Nevsky[3]**

[1] Southern Federal University (SFU), Rostov-on-Don, Russian Federation;
E-mails: larisavodol@aaatec.org, larisavodol@yahoo.com
[2] Donetsk Regional Museum, Donetsk, Ukraine; E-Mail: doold@mail.ru
[3] Southern Federal University (SFU), Rostov-on-Don, Russian Federation; E-mail: munevsky@sfedu.ru

**Abstract**

The article presents the results of multidisciplinary study carried out with the help of archaeological and astronomical methods. The aim of the study was to analyze and interpret the signs - elements of the composition, incised on the outer side surface of the vessel of the Late Bronze Age, owned to Srubna culture and discovered near the Staropetrovsky village in the northeast of the Donetsk region. The measurements and astronomical calculations revealed that all signs have astronomical meaning. Fourray star has been interpreted as the star Sirius. The sign polyline has been interpreted as an analog of graphic of equation of time in which the testimony of a water clock correspond with average solar time, and the testimony of a sundial - the true solar time. Sign wheel has been interpreted as a complex of lines, reflecting the regularity changing the direction of the shadows from the gnomon at sunset at the equinoxes and solstices. The sign thin polyline has been interpreted as a symbol of change of height of the Sun in the meridian at the equinoxes and solstices. Interpretation of wheel and polyline is related to the definition of the duration of the day. Sign of the wheel shows a direct relationship the duration of the day on the magnitude of the azimuth of sunrise/sunset, and thin polyline - from changing the height of the Sun in the meridian. Thus Staropetrovsky vessel, thanks to the astronomical signs, is a kind of ancient astronomical visual aid, which reflects the level of astronomical knowledge of the priesthood of Srubna culture at the late Bronze Age.

**Analysis and interpretation of the signs "star" and "polyline"**

In 1985, near the Staropetrovsky village ($48^0 13'$ N, $38^0 09'$ E) a clay pot, owned Srubna culture and dating XV-XIV centuries BC, was found in a ruined barrow [1, p. 102, 107–108], [2], [3, p. 89, 91]. Its uniqueness is the simultaneous presence of marks on its inner surface, which is in itself unique, and the marks on the outer surface of the vessel. To analyze the marks Staropetrovsky vessel, a series of interdisciplinary research by means of complex natural science methods, which have already demonstrated their effectiveness in these studies was performed [4, p. 5-13], [5-12]. Unique marks on its inner surface turned marking of a water clock [13], and the wells on the aureole fulfilled the function of hour markers of the horizontal sundial with sloping gnomon [14]. Starlike marks on the aureole of the Staropetrovsky vessel noted time of heliacal rising of Sirius on the sundial "dial" [14].

Complicated composition of schematic incised marks made on the outer side of surface of Staropetrovsky vessel. For today there are more than 300 vessels with signs of Srubna culture [3]. It should be noted that for the Srubna marks in general is characteristic the schematic character of images at which an informative function predominates over artistic.

Sign of the "star" - the first element of the composition on the outer side surface of the vessel, which resembles a four-star, located directly under the starlike marks of the corolla (Fig. 1). We assumed that the sign "star" symbolizes Sirius, and its location at the beginning of the composition symbolizes the beginning of astronomical year.

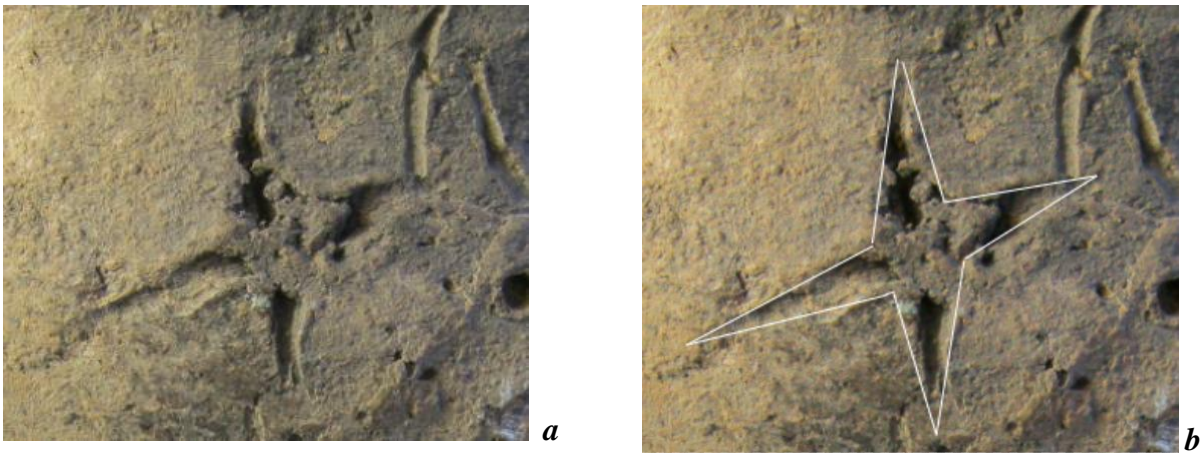

**Figure 1.** Vil. Staropetrovsky, ruined barrow, vessel with the marks on the inner surface: *a* – photo star from composition; *b* – schematic drawing of a star.

Initially, the assumption about the use staropetrovsky vessel as water clocks, arose in connection with the need to count the control periods of time lasting one hour at the markup analemmatic sundial [12]. However, sundials always show local true solar time, and water - the mean solar time. The exact duration of the true solar 24-hour day vacillates, so reading a sundial and water clock do not always coincide. From the point of view of modern astronomy, the true solar day - is the time interval between two successive upper or lower culminations[1] of the Sun on the same geographical meridian [15, p. 33-35]. The unevenness of diurnal movement of the Sun caused the ellipticity of the Earth's orbit around the Sun and the tilt of Earth's axis to the ecliptic plane. The duration of the mean solar day is the average duration of the true solar 24-hour days in a year. The difference between the true solar time and mean time in the same moment is called the equation of time[2]:

---

[1] The phenomenon of crossing celestial object celestial meridian - a great circle of the celestial sphere, plane which passes through the plumb line and the celestial axis, called the culmination of celestial object.
[2] Sometimes used reversed equation of time equal to the difference between the mean time and true solar time.

$$\eta = T_s - T_m \ , \tag{1}$$

where $\eta$ - the equation of time, $T_s$ - true solar time, $T_m$ - mean solar time.

We have calculated the value of the equation of time for the 1400 BC (see. Application) and built its graph (Fig. 2). The origin of coordinates corresponds to the beginning of the astronomical year - the date of heliacal rising of Sirius. We calculated it using astronomical program RedShift-7 Advanced. At the latitude of detection Staropetrovsky vessel heliacal rising of Sirius in 1400 BC occurred August 9 - later 34 days after the summer solstice (for the modern era corresponds to July 25, 2015 - see. Application) and observed on August 9 in the range of ± 100 years, approximately. The calculations were performed using an astronomical computer program HORIZONS System [3].

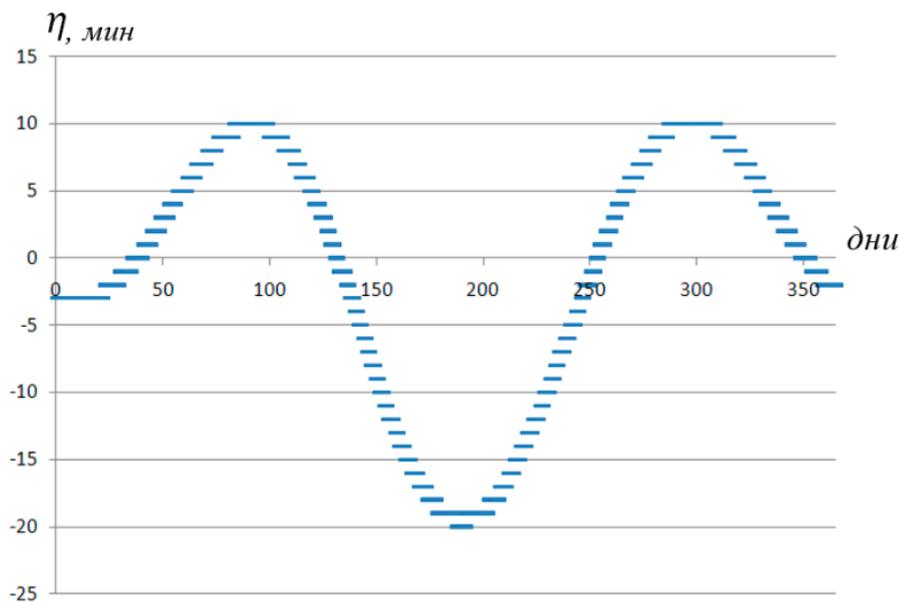

**Figure 10.** Graph the equation of time: $\eta$ - the equation of time. On the ordinate - the value of the equation of time in minutes, on the abscissa - number of days from the beginning of the year.

Polyline of the composition, by its shape resembles the graph of the equation of time (Fig. 11). Wherein the horizontal axis corresponds to null deviation from the mean solar time. Three point indentations under the polyline are arranged so that each indentation can be associated with one of the extremes (peaks) in the chart.

Extreme indentations located below the central and correspond to peaks polyline when readings the water clock will lag behind the readings of a sundial, i.e. water level in the water clock will be lower, than on the sundial at the same time, but on other days.

Central indentation is, respectively, above the central and corresponds to the peak when the water clock readings will outstrip readings of a sundial, ie the water level in the water clock will be higher than on the sundial at the same time, but on other days. I.e. these indentations could show peaks inverted graph of the equation of time and play the role of small tips for the proper use of water clocks.

---

[3] http://ssd.jpl.nasa.gov/?horizons

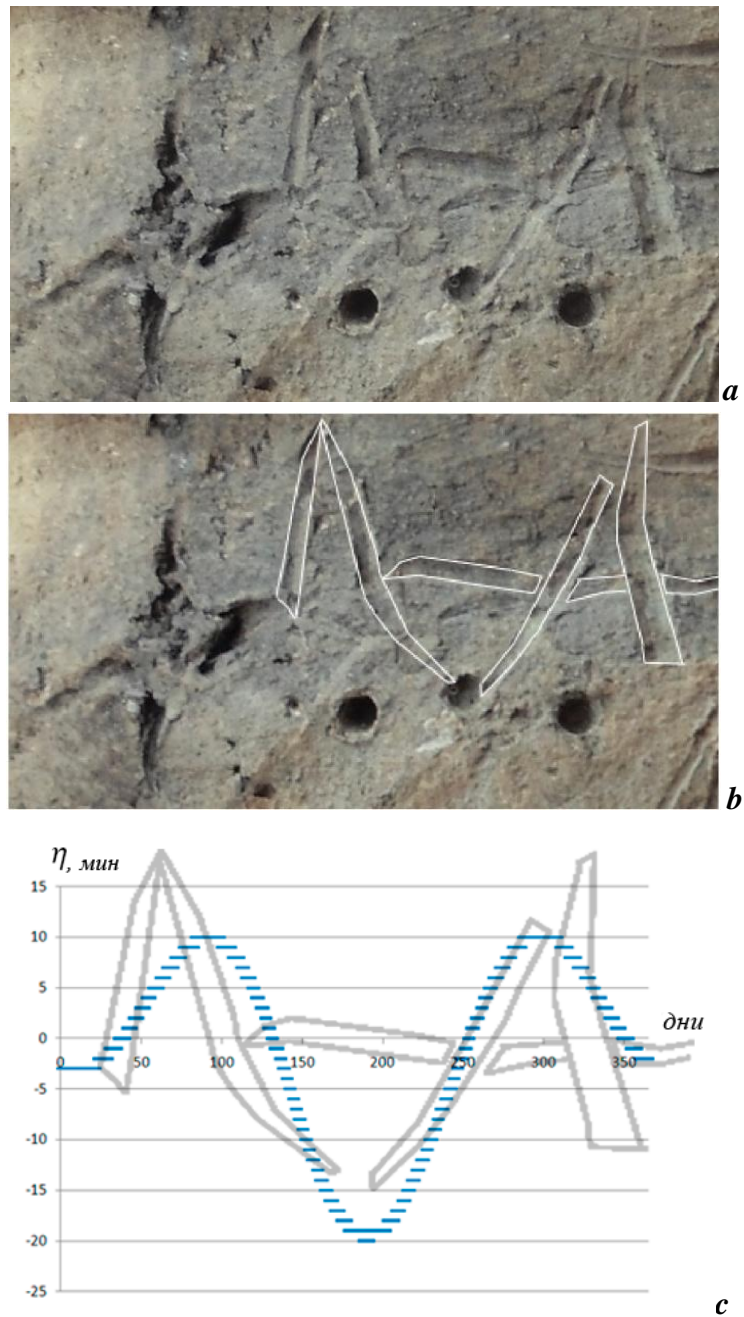

**Figure 3.** Vil. Staropetrovsky, ruined barrow, vessel with the marks on the inner surface, *a* - photo composition of polyline; *b* - schematic drawing of polyline, *c* - combination of polyline drawing and graphics of equation of time (see. fig. 10).

In Babylon, in the II century BC already knew about the deviation of the true solar time from mean solar. At the same time there were sufficiently well and in detail, developed a theory describing the uneven apparent motion of the Sun [16, p. 6 -7], [17, p. 213–215]. Ptolemy - Late Hellenistic astronomer from Alexandria (Egypt) - in his fundamental work "Almagest" (Book III, Part 9) also considered the issue of uneven movement of the Sun [18, p. 171].

However, all of these theories could not appear suddenly. For their formation took years of systematic observations of the motion of the Sun. We believe that the fragment composition polyline on staropetrovsky vessel, just, and is one of such evidence observations in earlier epochs of non-uniformity of the annual motion of the Sun.

For comparison we have built graphs of the equation of time with the origin of coordinates coinciding with the days of the solstices and equinoxes as well as for the geographical coordinates of Memphis (Fig. 4).

For Memphis (29°50′40″ N, 31°15′03″ E) we calculated the date of heliacal rising of Sirius (Sirius height ≈+3°, at the height of the Sun ≈-6°), 17.07.1400 BC - the 11-th day after the summer solstice (for the modern age corresponds to July 2, 2015 regarding the date of the summer solstice).

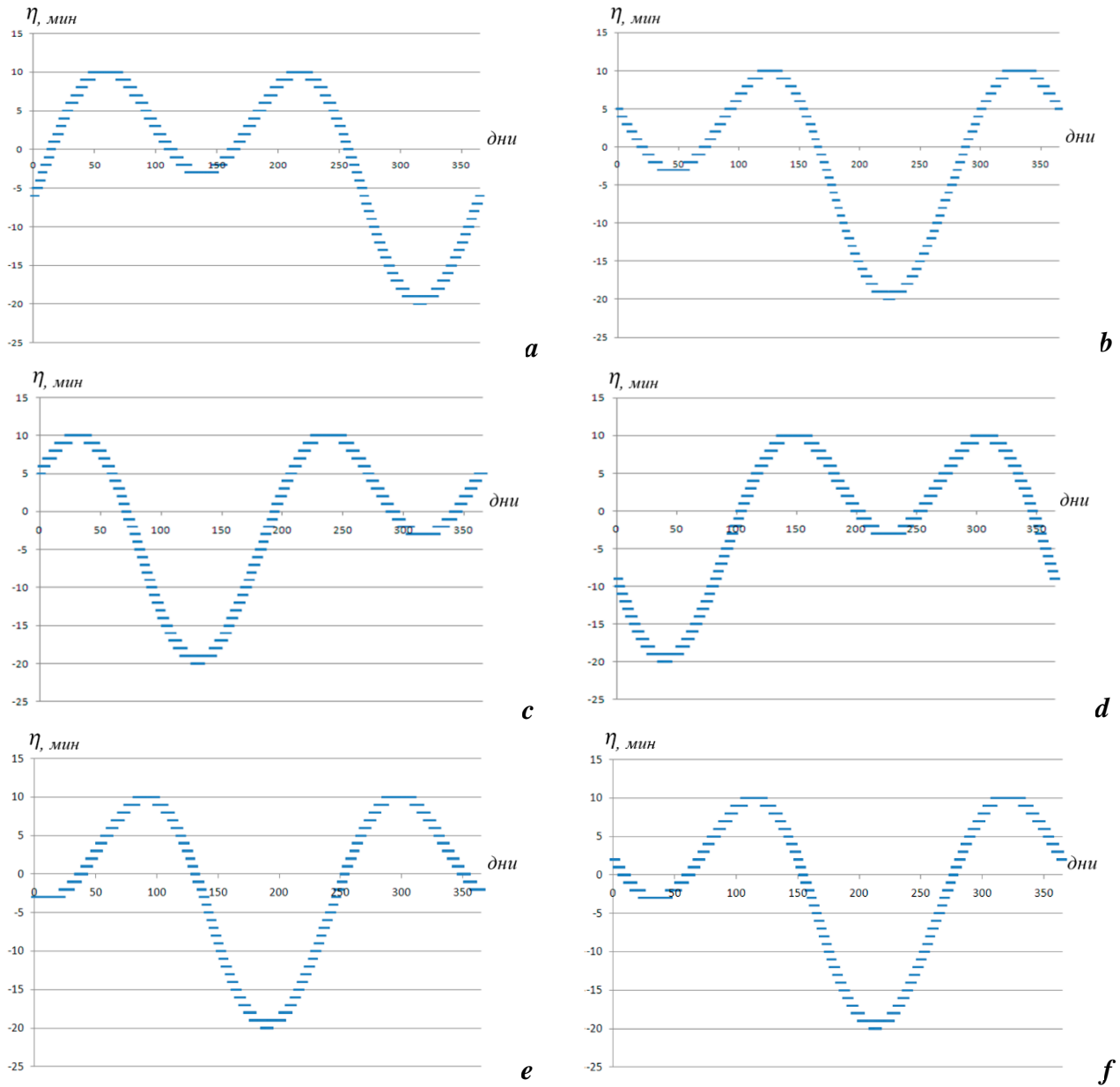

**Figure 4.** Graph of the equation of time for the beginning of the year according to the date: a - the vernal equinox, b - the summer solstice, c - the autumnal equinox, d - the winter solstice, e - heliacal rising of Sirius at the latitude of the village. Staropetrovsky, f - heliacal rising of Sirius at the latitude of Memphis. On the ordinate - the value of the equation of time in minutes, on the abscissa - number of days from the beginning of the year.

Charts equation of time for heliacal rising of Sirius in Memphis (Fig. 4e) and for latitude of village staropetrovsky (Fig. 4f) are very similar. However, the value of the equation of time at the end of the year takes a negative value – line of graph intersects the horizontal axis similarly polyline in the composition of staropetrovsky vessel not for Memphis, but only for the village Staropetrovsky. Thus, if polyline really is a schematic representation of the graph of the equation of time, then it is the image of graphic for latitude close to the latitude of the village. Staropetrovsky, rather than for the more southern latitudes.

In order to make sure that the accuracy of the dating of the vessel will not greatly affect on the shape of the graph of the equation of time and, consequently, on the interpretation of the polyline, we have carried out additional calculations of the equation of time for the 1900 BC and 900 BC. Calculations showed that, graphs of equations of time for 1400 BC ± 500 years, built regarding the date of heliacal rising of Sirius on the latitude of the Staropetrovsky village, visually almost indistinguishable, especially given the very schematic representation of the polyline on the surface of the vessel (Fig. 5).

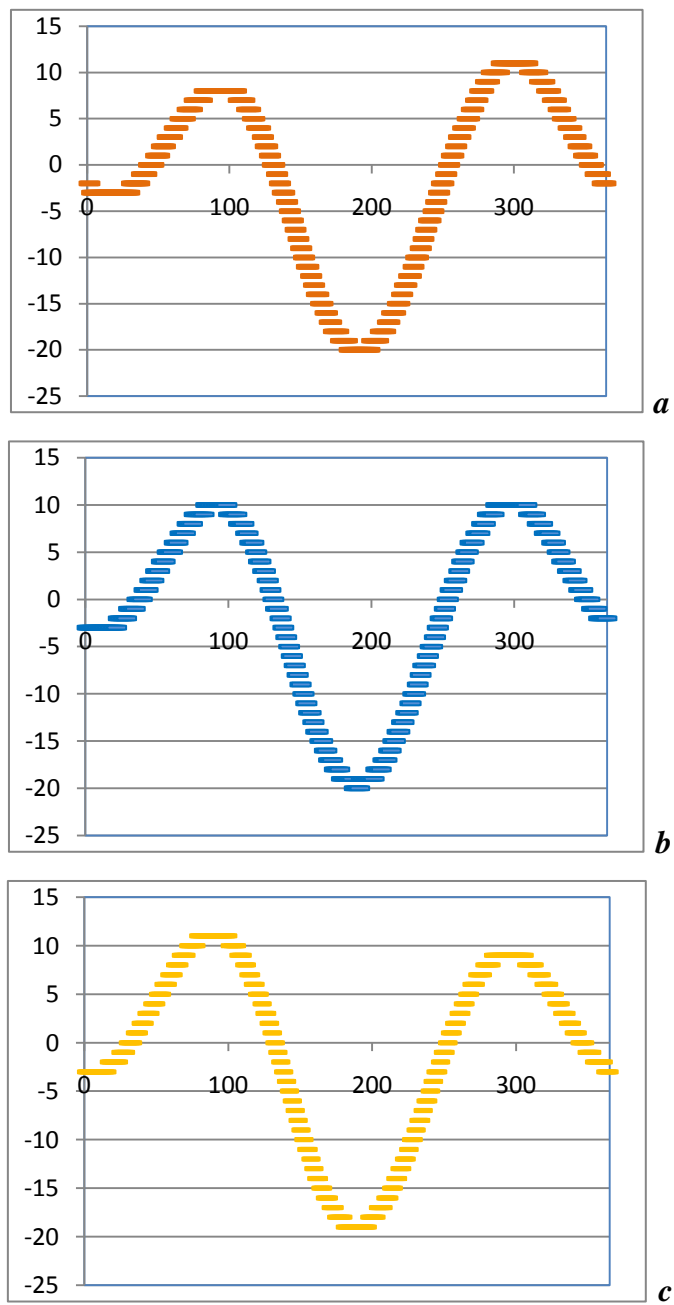

**Figure 5.** The graph of the equation of time for the beginning of the year, corresponding to the date of heliacal rising of Sirius: *a* - 1900 BC, *b* - 1400 BC, *c* - 900 BC.

Thus, the observed visual similarity we can consider as evidence in favor of the calendar system close to the ancient Egyptian, with the beginning of the year, measured from the date of heliacal rising of Sirius, at the Srubna population of Northern Black Sea Coast.

**Analysis and interpretation of the sign "wheel"**

The next sign after the polyline is a central sign of the whole composition on the outer surface of the vessel Staropetrovsky - "wheel" (Fig. 6). I.e. circular signs in decorative patterns interpret often, as solar, we supposed, what the "wheel" of the composition could be connected in meaning with the Sun. In connection with this assumption, we tried to compare with the "wheel spokes" azimuths of sunrise and sunset and the azimuths of the shadow, which arises from the gnomon at sunrise and sunset, at the equinoxes and solstices.

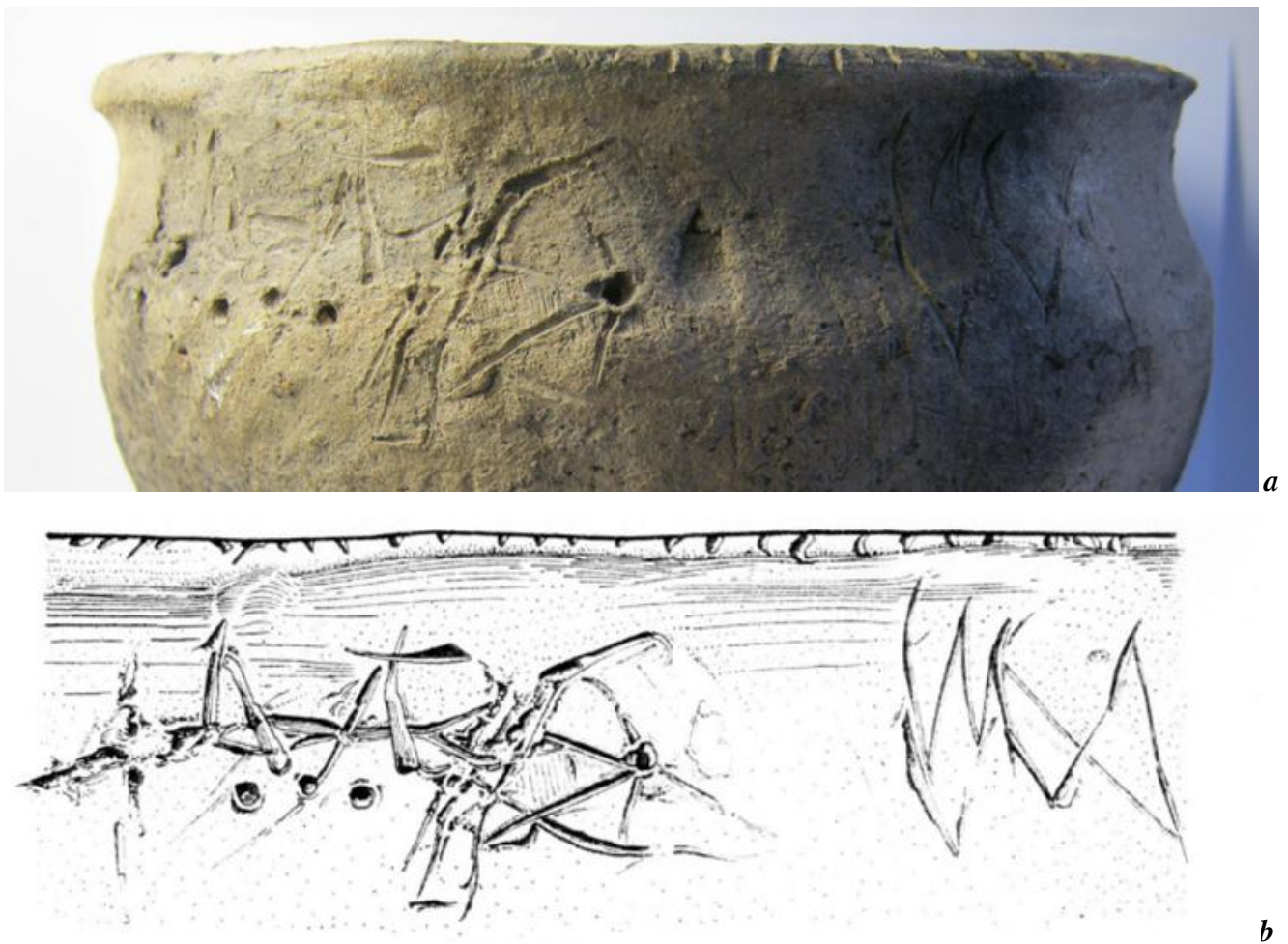

*a*

*b*

**Figure 6.** Vil. Staropetrovsky, ruined barrow, vessel with the marks on the inner surface: *a* – photo of the composition on the outside of the vessel (photo by A.N. Usachuk, 2014), *b* - drawing of the composition (drawing by V.B. Pankovsky and A.N. Usachuk, 1993).

The azimuths of sunrise in days equinoxes and solstices were expected on formula 2 [19, p. 21], and azimuths of sunset and azimuths of shadow from the gnomon on formulas 3-5:

$$\cos A_r = \frac{\sin \delta - \sin \varphi \cdot \sin h}{\cos \varphi \cdot \cos h}, \tag{2}$$

$$A_s = 360^0 - A_r, \tag{3}$$

$$A_{shr} = A_r + 180^0, \tag{4}$$

$$A_{shs} = A_s - 180^0, \tag{5}$$

where $A_r$ - azimuth of sunrise, a read-out from the North to East (geodesic), $A_s$ - azimuth of setting of Sun, $A_{shr}$ – the azimuth of the shadow at sunrise, $A_{shs}$ - azimuth of the shadow at sunset, $\delta$ - declension of the Sun, $h$ - height of the Sun above horizon, $\varphi$ - geographical latitude. The height of the Sun in the moment of rising (setting) calculated for the overhead edge of Sun disk:

$$h = -R - \rho + p, \tag{6}$$

where $R$ - angular radius of the Sun, $\rho$ - refraction at the horizon, $p$ - horizontal parallax [20, p. 19]. For the Sun $R=16'$, $\rho=35'$ [21, p. 44], $p=8.8''$ [21, p. 36].

During summer solstice the Sun declination equal to angle of ecliptic inclination to celestial equator $\varepsilon$, which is calculated using the formula [22, p. 35]:

$$\varepsilon = 23.43929111^0 - 46.8150'' \cdot T - 0.00059'' \cdot T^2 + 0.001813 \cdot T^3, \tag{7}$$

$$T \approx \frac{(y - 2000)}{100}, \tag{8}$$

where $T$ - the number of Julian centuries, that separates this age from noon of the 1 of January 2000, $y$ - year of required age. During winter solstice the Sun declination $\delta = -\varepsilon$, and during equinoxes $\delta = 0$.

Calculated by us according to the formula 7 the angle of inclination of the ecliptic to the celestial equator $\varepsilon=23^051'41''$ for 1400 BC. The most interesting results appeared azimuths of the shadow directions in the days of equinoxes and solstices at sunset calculated by the formulas 4-5. The results of our calculations are shown in Table 1.

**Table 10.** Azimuths of the Sun at sunrise (sunset) at the equinoxes and solstices, calculated on the top edge of the visible disk of the Sun; $h$ - height of the Sun, $\delta$ - declination of the Sun, $A_r$ - azimuth of the Sun, $A_{shr}$ – the azimuth of the shadow at sunrise, $A_{shs}$ - azimuth of the shadow at sunset.

| phenomenon | $h$, ($^0$) | $\delta$, ($^0$) | $A$, ($^0$) | $A_{shs}$, ($^0$) | $A_{shr}$, ($^0$) |
|---|---|---|---|---|---|
| summer solstice, sunrise | -0,85 | 23,86 | 51,41 | - | 231,41 |
| equinoxe, sunrise | -0,85 | 0,00 | 89,05 | - | 269,05 |
| winter solstice, sunrise | -0,85 | -23,86 | 126,20 | - | 306,20 |
| summer solstice, sunset | -0,85 | 23,86 | 308,59 | 128,59 | - |
| equinoxe, sunset | -0,85 | 0,00 | 270,95 | 90,95 | - |
| winter solstice, sunset | -0,85 | -23,86 | 233,80 | 53,80 | - |

In the center of "wheel" there is deepening in that it is possible to stick a thin rod, for example, straw as a gnomon that would throw shadow at illumination a Sun (fig. 7).

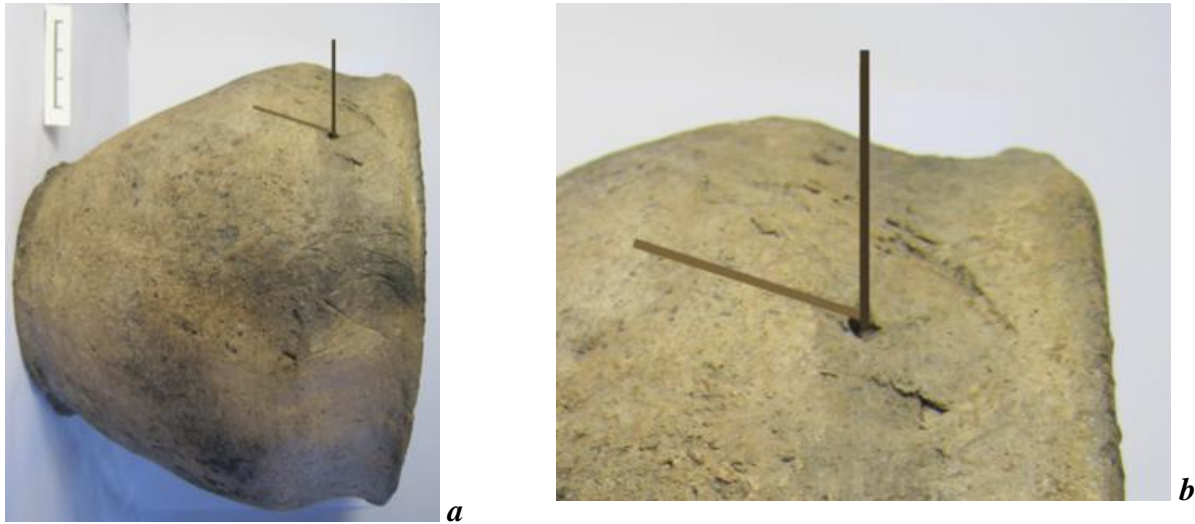

**Figure 7.** Vil. Staropetrovsky, ruined barrow, vessel with the marks on the inner surface: modeling of the gnomon in the center of "wheel": ***a*** - kind of the whole the vessel with gnomon, ***b*** - a fragment of the vessel with a "wheel" with gnomon.

Within the framework of one of our hypotheses, if to turn a standing vessel under the corner of 90° ″by a wheel″ upwards, then a lower (for a standing vessel) ″spoke″ will correspond to shade, by cast aside such straw at midday, and, accordingly, to specify direction on North. Thus direction of the first spoke, if to consider clockwise from a lower ″spoke″, will approximately coincide with direction of shade that is cast aside by a straw at setting of Sun in the day of winter solstice. Direction of the second spoke will coincide with direction of shade at setting of Sun in the day of equinox, and third spoke - at setting of Sun in the day of summer solstice (fig. 8). The coincidence of directions was examined by us, as approximate, as a surface of vessel in area of ″wheel″ is distorted, that does not allow correctly calculate exactness of such coincidence.

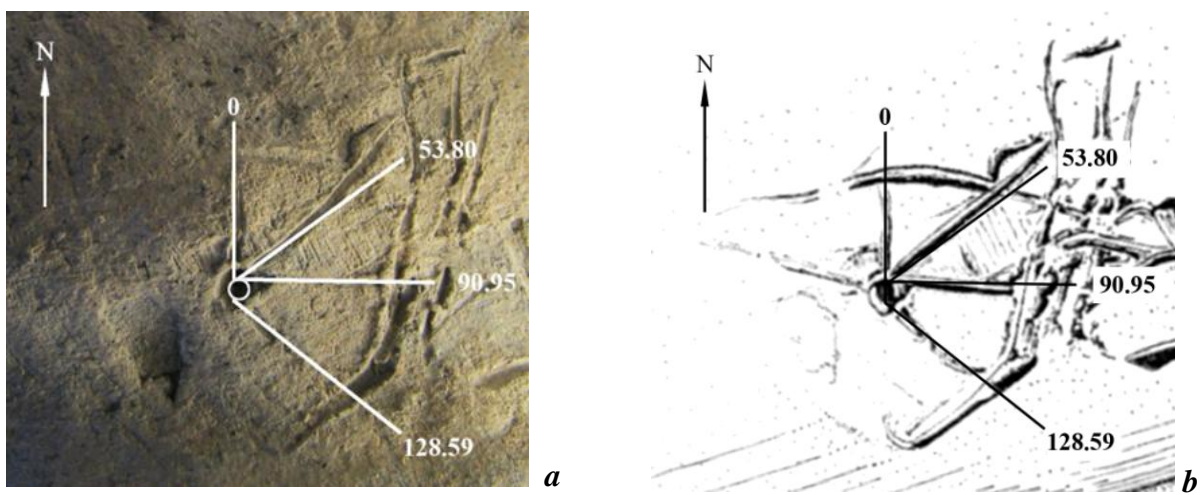

**Figure 8.** Vil. Staropetrovsky, ruined barrow, vessel with the marks on the inner surface: ″wheel″ from composition with the inflicted azimuths of direction of shade, corresponding to setting of Sun in days an equinox, summer and winter solstice: ***a*** - photo, ***b*** - drawing. *N* - True North.

Parts of the spokes adjacent to the center of the wheel coincided with the direction of the shadow most accurately. Since "spokes", as mentioned above, when patterning is drawn from the central depressions to a "rim" wheels, therefore, adjacent to the center part of the "spokes" were drawn more closely and their directions were more significant.

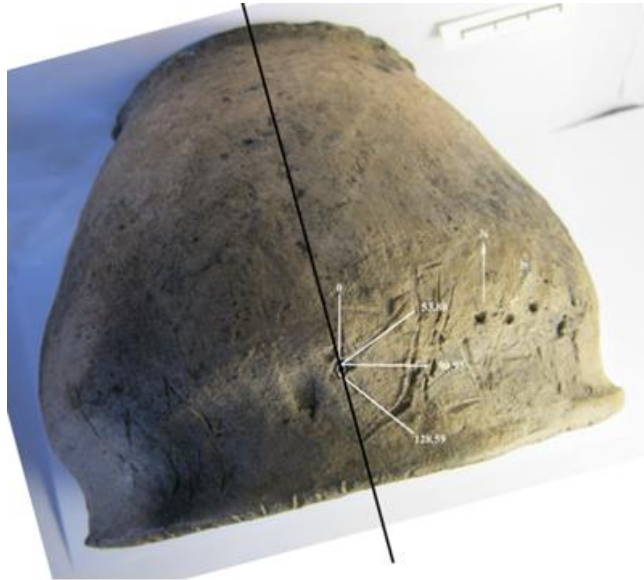

**Figure 9.** Vil. Staropetrovsky, ruined barrow, vessel with the marks on the inner surface, deviation the vessel (the vertical axis of symmetry of the vessel) on the direction to the North "wheel". The symmetry axis is applied the black line. *N* - direction of True North.

At repeated studies composition Staropetrovsky vessel was found that the character of making yet another - the fifth "spokes" are qualitatively different from character making the first four "spokes" (Fig. 6). Perhaps the line, which was originally adopted for the "spoke" is a casual track arisen during processing (smoothing down) the outer surface of the vessel. Therefore, its spatial arrangement in the analysis of "wheel" are not taken into account.

A lower wheel arm indicative on North is declined from the vertical axis of symmetry of vessel on a corner approximately equal 15º (fig. 9). It a size corresponds to the corner to that a Sun is moved on an ecliptic for one hour (360º for 24 hours) and can be used for approximate determination of duration of day.

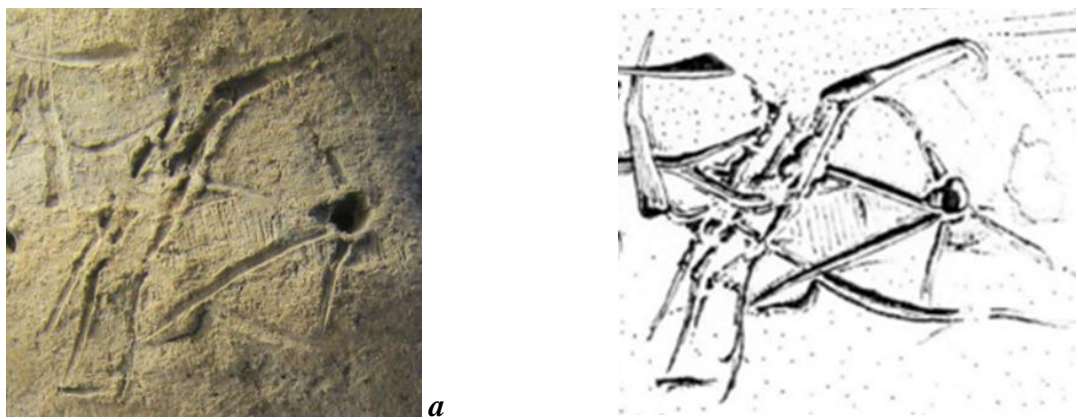

*a*                                                                                           *b*

**Figure 10.** Vil. Staropetrovsky, ruined barrow, vessel with the marks on the inner surface: *a* – photo drawing of the wheel; *b* - schematic drawing of the wheel.

Another sign in the composition, which in our view, may be associated with the calendar calculating time - three thin line at the bottom left near the wheel (Fig. 10).

Wheel in the composition on Staropetrovsky vessel semantic and operatively associated with the Sun and can symbolize the solar year[4] - the length of time for which the Sun completes one cycle of changing seasons, for example, from one day of the summer solstice, to the other. In the tropical year contains approximately 365.24 days. Interval between the two heliacal rising of Sirius is about 365.25 days. These values are close, and perhaps originally, the duration of a year of 365 days was determined by the Egyptians it was thanks to the observations of the heliacal rising of Sirius. The number of whole days between its heliacal rising, is 365 days, but only for a period of 3 years. In the fourth year, between its heliacal rising, will be 366 days, then all will be repeated. Apparently so the Egyptians have officially adopted the length of the year equal to 365 days. For every 4 years seasonal phenomena outpaced calendar on 1 day and there is a the need for introducing a leap year, which was formally introduced in Egypt only in the III century BC. When the systematic observation of the heliacal rising stars was easy to notice the regularity associated with repetition of 365 days - period for three years. We therefore believe that the three lines next to the drawing wheel can denote these three repeating, when solar annual "wheel", "rotates on one place" as though.

**Analysis and interpretation of the sign "thin polyline"**

The second group of characters, located to the right of "wheels" and represents another polyline applied thinner lines - "thin polyline " (Fig. 11).

The second group of characters - "thin polyline " - consists of three peaks, and, in our view, could symbolize change the maximum height of the sun in the upper culmination (at noon) and changes of the angular distance between the azimuths of sunrise and sunset during the winter solstice (the peak with the apex *A1*), equinox (peak apex *B1*) and the summer solstice (the peak with the top of *C1*). The height of the Sun at the time of upper culmination and time of sunrise and sunset for the days of solstices and equinoxes are shown in Figure 12.

The value of the duration of the day from sunrise to sunset at the summer solstice $D_{ss}$ refers to the duration of the day in the equinox $D_{eq}$ and duration of the day at the winter solstice $D_{eq}$, for the latitude detection staropetrovsky vessel as $D_{ss}:D_{eq}:D_{ws}$=4:3:2=2:1.5:1 [13].

To make it easier to analyze the height of the peaks, we turned on the drawing image "thin polyline " so that the vertices *A1, B1, C1* are aligned (Fig. 15 b), and measure the height of the peaks. The height of the peak *C1*, we examined how the projection of the segment [*C1C2*] on the vertical axis, $c \approx 1.5$ cm. In this case, is clearly seen that the segments [*D1D2*] and [*E1E2*] have approximately the same length and inclination as the segment [*C1C2*]. Very likely that intervals [D1D2] and [E1E2] were standards of length when applying "thin polyline" and carrying out functions samples, based on which were drawn the other lines.

The height of the peak *B1*, we examined how the projection of the segment [*B1B2*] on the vertical axis, $b \approx 2.2$ cm. Strongly curved line of the segment [*B1B2*], most likely a random defect. Point *B2* and *E3* are approximately at the same level, so perhaps segment [*E2E3*] has a supporting role to determine the location of the point *B2*.

The height of the peak *A1* has been considered by us as the projection of the segment [*A1A2*] on the vertical axis. However, bearing in mind that the application of the line [*A1A2*], most likely, as the rest of the lines was based on samples [*D1D2*] and [*E1E2*], and the fact that the segment [*A1A2*] ends surface defect, we placed point *A2* in the area of defect on the same level as the point *E2* on segment-benchmark [*E1E2*]. The height of the peak *A1*, wherein, equal *a*≈2.9 cm.

---

[4] Such solar year in modern astronomy is called a tropical

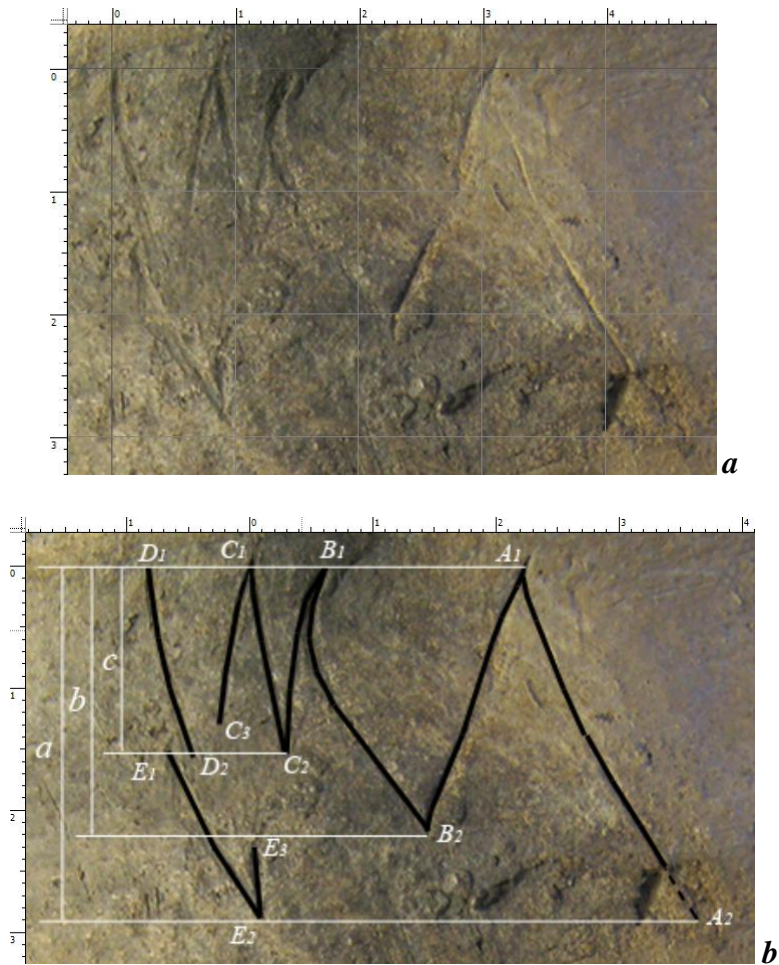

**Figure 11.** Vil. Staropetrovsky, ruined barrow, vessel with the marks on the inner surface, drawing "thin polyline": ***a*** – photo of the drawing; ***b*** - schematic drawing of a "thin polyline ".

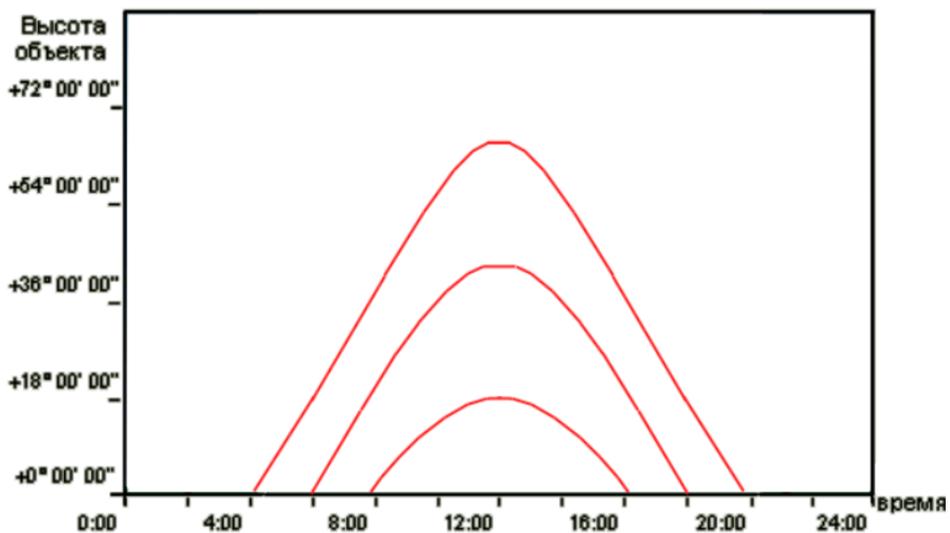

**Figure 12.** Graph of the height of the Sun depending on the time of day in the days of winter solstice (lower graph), equinox (middle graph) and the summer solstice (upper graph). On the ordinate - height, along the horizontal axis - time.

The ratio of the height of the peaks $a:b:c \approx 2.9:2.2:1.5=1.9:1.5:1$ is very close to the ratio of the length of the day for the solstices and equinoxes $D_{ss}:D_{eq}:D_{ws}=4:3:2=2:1.5:1$, and this confirms our hypothesis that the "thin polyline symbolizes change the duration of the day throughout the year.

**Conclusion**

Thus, in the process of studying of Staropetrovsky vessel were fully analyzed and interpreted all signs on its outer side surface. The analysis revealed that all signs have astronomical meaning. Fourray star has been interpreted as the brightest star in the night sky - Sirius.

Sign polyline, which is located right after the star, has been interpreted as an analog of graphic of equation of time in which the testimony of a water clock correspond with average solar time, and the testimony of a sundial - with a true solar time throughout the year. Start of schedule, as well as the beginning of the year, was related to the date of heliacal rising of Sirius at Late Bronze Age for the latitude of the Staropetrovsky village.

Following the polyline, a sign of "Wheel" has been interpreted as a complex of lines, reflecting the regularity changing the direction of the shadows from the gnomon, at sunset at the equinoxes and solstices. Using it could be approximately determine the change in length of the day throughout the year, depending on the azimuth of sunrise / sunset.

The sign thin polyline has been interpreted as a symbol of change in height of the Sun, in the meridian at the equinoxes and solstices. The ratio of the peak heights of thin polyline fairly accurately reflects the ratio of the length of the day from sunrise to sunset in these days.

Interpretation of both signs - wheel and thin polyline - are related to the definition of the duration of the day. Sign wheel shows a direct correlation on the value of the duration of the day azimuth of sunrise/sunset, and thin polyline - changing the height of Sun on the meridian.

Spite of the fact that Staropetrovsky vessel is marked with guides and water clocks, and sundials, Staropetrovsky vessel could not be used for direct determine the time of heliacal rising of Sirius, because price of division of water clock "scale" is one hour and with its help it is impossible to measure the time to the nearest minute.

For such measurements should have existed more accurate clepsydras. Staropetrovsky vessel served for a rough determination of the time with accurate to one hour, required, for example, in everyday life or in religious rituals, but not in the exact astronomical observations. Perhaps Staropetrovsky vessel also functions as astronomical visual aid, which, thanks to the semantic content of the sign on the outer side surface, reflects the level of astronomical knowledge of the priesthood of Srubna culture at the late Bronze Age.

**Acknowledgement**

The authors express their sincere thanks to astronomer of European Southern Observatory Nando Patat for the consultations on using the program HORIZONS System.

Application

| Date 1400 BC | Date 2015 AD | η, (min) | Date 1400 BC | Date 2015 AD | η, (min) | Date 1400 BC | Date 2015 AD | η, (min) |
|---|---|---|---|---|---|---|---|---|
| 1.1 | 17.12 | -7 | 12.2 | 28.1 | -20 | 26.3 | 11.3 | -10 |
| 2.1 | 18.12 | -7 | 13.2 | 29.1 | -20 | 27.3 | 12.3 | -10 |
| 3.1 | 19.12 | -8 | 14.2 | 30.1 | -20 | 28.3 | 13.3 | -9 |
| 4.1 | 20.12 | -8 | 15.2 | 31.1 | -20 | 29.3 | 14.3 | -9 |
| 5.1 | 21.12 | -9 | 16.2 | 1.2 | -20 | 30.3 | 15.3 | -8 |
| 6.1 | 22.12 | -9 | 17.2 | 2.2 | -19 | 31.3 | 16.3 | -8 |
| 7.1 | 23.12 | -10 | 18.2 | 3.2 | -19 | 1.4 | 17.3 | -7 |
| 8.1 | 24.12 | -10 | 19.2 | 4.2 | -19 | 2.4 | 18.3 | -7 |
| 9.1 | 25.12 | -11 | 20.2 | 5.2 | -19 | 3.4 | 19.3 | -7 |
| 10.1 | 26.12 | -11 | 21.2 | 6.2 | -19 | 4.4 | 20.3 | -6 |
| 11.1 | 27.12 | -12 | 22.2 | 7.2 | -19 | 5.4 | 21.3 | -6 |
| 12.1 | 28.12 | -12 | 23.2 | 8.2 | -19 | 6.4 | 22.3 | -5 |
| 13.1 | 29.12 | -12 | 24.2 | 9.2 | -19 | 7.4 | 23.3 | -5 |
| 14.1 | 30.12 | -13 | 25.2 | 10.2 | -19 | 8.4 | 24.3 | -5 |
| 15.1 | 31.12 | -13 | 26.2 | 11.2 | -19 | 9.4 | 25.3 | -4 |
| 16.1 | 1.1 | -14 | 27.2 | 12.2 | -18 | 10.4 | 26.3 | -4 |
| 17.1 | 2.1 | -14 | 28.2 | 13.2 | -18 | 11.4 | 27.3 | -3 |
| 18.1 | 3.1 | -14 | 1.3 | 14.2 | -18 | 12.4 | 28.3 | -3 |
| 19.1 | 4.1 | -15 | 2.3 | 15.2 | -18 | 13.4 | 29.3 | -2 |
| 20.1 | 5.1 | -15 | 3.3 | 16.2 | -18 | 14.4 | 30.3 | -2 |
| 21.1 | 6.1 | -15 | 4.3 | 17.2 | -17 | 15.4 | 31.3 | -2 |
| 22.1 | 7.1 | -16 | 5.3 | 18.2 | -17 | 16.4 | 1.4 | -1 |
| 23.1 | 8.1 | -16 | 6.3 | 19.2 | -17 | 17.4 | 2.4 | -1 |
| 24.1 | 9.1 | -16 | 7.3 | 20.2 | -17 | 18.4 | 3.4 | 0 |
| 25.1 | 10.1 | -17 | 8.3 | 21.2 | -16 | 19.4 | 4.4 | 0 |
| 26.1 | 11.1 | -17 | 9.3 | 22.2 | -16 | 20.4 | 5.4 | 1 |
| 27.1 | 12.1 | -17 | 10.3 | 23.2 | -16 | 21.4 | 6.4 | 1 |
| 28.1 | 13.1 | -17 | 11.3 | 24.2 | -15 | 22.4 | 7.4 | 1 |
| 29.1 | 14.1 | -18 | 12.3 | 25.2 | -15 | 23.4 | 8.4 | 2 |
| 30.1 | 15.1 | -18 | 13.3 | 26.2 | -15 | 24.4 | 9.4 | 2 |
| 31.1 | 16.1 | -18 | 14.3 | 27.2 | -14 | 25.4 | 10.4 | 2 |
| 1.2 | 17.1 | -18 | 15.3 | 28.2 | -14 | 26.4 | 11.4 | 3 |
| 2.2 | 18.1 | -18 | 16.3 | 1.3 | -14 | 27.4 | 12.4 | 3 |
| 3.2 | 19.1 | -19 | 17.3 | 2.3 | -13 | 28.4 | 13.4 | 4 |
| 4.2 | 20.1 | -19 | 18.3 | 3.3 | -13 | 29.4 | 14.4 | 4 |
| 5.2 | 21.1 | -19 | 19.3 | 4.3 | -13 | 30.4 | 15.4 | 4 |
| 6.2 | 22.1 | -19 | 20.3 | 5.3 | -12 | 1.5 | 16.4 | 5 |
| 7.2 | 23.1 | -19 | 21.3 | 6.3 | -12 | 2.5 | 17.4 | 5 |
| 8.2 | 24.1 | -19 | 22.3 | 7.3 | -12 | 3.5 | 18.4 | 5 |
| 9.2 | 25.1 | -19 | 23.3 | 8.3 | -11 | 4.5 | 19.4 | 6 |
| 10.2 | 26.1 | -19 | 24.3 | 9.3 | -11 | 5.5 | 20.4 | 6 |
| 11.2 | 27.1 | -19 | 25.3 | 10.3 | -10 | 6.5 | 21.4 | 6 |

| Date 1400 BC | Date 2015 AD | η, (min) | Date 1400 BC | Date 2015 AD | η, (min) | Date 1400 BC | Date 2015 AD | η, (min) |
|---|---|---|---|---|---|---|---|---|
| 7.5  | 22.4 | 6  | 20.6 | 5.6  | 8  | 3.8  | 19.7 | -2 |
| 8.5  | 23.4 | 7  | 21.6 | 6.6  | 8  | 4.8  | 20.7 | -2 |
| 9.5  | 24.4 | 7  | 22.6 | 7.6  | 8  | 5.8  | 21.7 | -2 |
| 10.5 | 25.4 | 7  | 23.6 | 8.6  | 8  | 6.8  | 22.7 | -2 |
| 11.5 | 26.4 | 7  | 24.6 | 9.6  | 8  | 7.8  | 23.7 | -2 |
| 12.5 | 27.4 | 8  | 25.6 | 10.6 | 7  | 8.8  | 24.7 | -2 |
| 13.5 | 28.4 | 8  | 26.6 | 11.6 | 7  | 9.8  | 25.7 | -3 |
| 14.5 | 29.4 | 8  | 27.6 | 12.6 | 7  | 10.8 | 26.7 | -3 |
| 15.5 | 30.4 | 8  | 28.6 | 13.6 | 7  | 11.8 | 27.7 | -3 |
| 16.5 | 1.5  | 9  | 29.6 | 14.6 | 7  | 12.8 | 28.7 | -3 |
| 17.5 | 2.5  | 9  | 30.6 | 15.6 | 6  | 13.8 | 29.7 | -3 |
| 18.5 | 3.5  | 9  | 1.7  | 16.6 | 6  | 14.8 | 30.7 | -3 |
| 19.5 | 4.5  | 9  | 2.7  | 17.6 | 6  | 15.8 | 31.7 | -3 |
| 20.5 | 5.5  | 9  | 3.7  | 18.6 | 6  | 16.8 | 1.8  | -3 |
| 21.5 | 6.5  | 9  | 4.7  | 19.6 | 5  | 17.8 | 2.8  | -3 |
| 22.5 | 7.5  | 10 | 5.7  | 20.6 | 5  | 18.8 | 3.8  | -3 |
| 23.5 | 8.5  | 10 | 6.7  | 21.6 | 5  | 19.8 | 4.8  | -3 |
| 24.5 | 9.5  | 10 | 7.7  | 22.6 | 4  | 20.8 | 5.8  | -3 |
| 25.5 | 10.5 | 10 | 8.7  | 23.6 | 4  | 21.8 | 6.8  | -3 |
| 26.5 | 11.5 | 10 | 9.7  | 24.6 | 4  | 22.8 | 7.8  | -3 |
| 27.5 | 12.5 | 10 | 10.7 | 25.6 | 4  | 23.8 | 8.8  | -3 |
| 28.5 | 13.5 | 10 | 11.7 | 26.6 | 3  | 24.8 | 9.8  | -3 |
| 29.5 | 14.5 | 10 | 12.7 | 27.6 | 3  | 25.8 | 10.8 | -3 |
| 30.5 | 15.5 | 10 | 13.7 | 28.6 | 3  | 26.8 | 11.8 | -3 |
| 31.5 | 16.5 | 10 | 14.7 | 29.6 | 3  | 27.8 | 12.8 | -3 |
| 1.6  | 17.5 | 10 | 15.7 | 30.6 | 2  | 28.8 | 13.8 | -3 |
| 2.6  | 18.5 | 10 | 16.7 | 1.7  | 2  | 29.8 | 14.8 | -3 |
| 3.6  | 19.5 | 10 | 17.7 | 2.7  | 2  | 30.8 | 15.8 | -3 |
| 4.6  | 20.5 | 10 | 18.7 | 3.7  | 2  | 31.8 | 16.8 | -2 |
| 5.6  | 21.5 | 10 | 19.7 | 4.7  | 1  | 1.9  | 17.8 | -2 |
| 6.6  | 22.5 | 10 | 20.7 | 5.7  | 1  | 2.9  | 18.8 | -2 |
| 7.6  | 23.5 | 10 | 21.7 | 6.7  | 1  | 3.9  | 19.8 | -2 |
| 8.6  | 24.5 | 10 | 22.7 | 7.7  | 1  | 4.9  | 20.8 | -2 |
| 9.6  | 25.5 | 10 | 23.7 | 8.7  | 0  | 5.9  | 21.8 | -2 |
| 10.6 | 26.5 | 10 | 24.7 | 9.7  | 0  | 6.9  | 22.8 | -2 |
| 11.6 | 27.5 | 10 | 25.7 | 10.7 | 0  | 7.9  | 23.8 | -1 |
| 12.6 | 28.5 | 10 | 26.7 | 11.7 | 0  | 8.9  | 24.8 | -1 |
| 13.6 | 29.5 | 10 | 27.7 | 12.7 | 0  | 9.9  | 25.8 | -1 |
| 14.6 | 30.5 | 9  | 28.7 | 13.7 | -1 | 10.9 | 26.8 | -1 |
| 15.6 | 31.5 | 9  | 29.7 | 14.7 | -1 | 11.9 | 27.8 | -1 |
| 16.6 | 1.6  | 9  | 30.7 | 15.7 | -1 | 12.9 | 28.8 | -1 |
| 17.6 | 2.6  | 9  | 31.7 | 16.7 | -1 | 13.9 | 29.8 | 0  |
| 18.6 | 3.6  | 9  | 1.8  | 17.7 | -1 | 14.9 | 30.8 | 0  |
| 19.6 | 4.6  | 9  | 2.8  | 18.7 | -2 | 15.9 | 31.8 | 0  |

| Date 1400 BC | Date 2015 AD | η, (min) | Date 1400 BC | Date 2015 AD | η, (min) | Date 1400 BC | Date 2015 AD | η, (min) |
|---|---|---|---|---|---|---|---|---|
| 16.9 | 1.9 | 0 | 25.10 | 10.10 | 9 | 3.12 | 18.11 | 6 |
| 17.9 | 2.9 | 0 | 26.10 | 11.10 | 9 | 4.12 | 19.11 | 6 |
| 18.9 | 3.9 | 1 | 27.10 | 12.10 | 9 | 5.12 | 20.11 | 5 |
| 19.9 | 4.9 | 1 | 28.10 | 13.10 | 9 | 6.12 | 21.11 | 5 |
| 20.9 | 5.9 | 1 | 29.10 | 14.10 | 9 | 7.12 | 22.11 | 4 |
| 21.9 | 6.9 | 1 | 30.10 | 15.10 | 9 | 8.12 | 23.11 | 4 |
| 22.9 | 7.9 | 2 | 31.10 | 16.10 | 10 | 9.12 | 24.11 | 4 |
| 23.9 | 8.9 | 2 | 1.11 | 17.10 | 10 | 10.12 | 25.11 | 3 |
| 24.9 | 9.9 | 2 | 2.11 | 18.10 | 10 | 11.12 | 26.11 | 3 |
| 25.9 | 10.9 | 2 | 3.11 | 19.10 | 10 | 12.12 | 27.11 | 3 |
| 26.9 | 11.9 | 3 | 4.11 | 20.10 | 10 | 13.12 | 28.11 | 2 |
| 27.9 | 12.9 | 3 | 5.11 | 21.10 | 10 | 14.12 | 29.11 | 2 |
| 28.9 | 13.9 | 3 | 6.11 | 22.10 | 10 | 15.12 | 30.11 | 1 |
| 29.9 | 14.9 | 3 | 7.11 | 23.10 | 10 | 16.12 | 1.12 | 1 |
| 30.9 | 15.9 | 4 | 8.11 | 24.10 | 10 | 17.12 | 2.12 | 0 |
| 1.10 | 16.9 | 4 | 9.11 | 25.10 | 10 | 18.12 | 3.12 | 0 |
| 2.10 | 17.9 | 4 | 10.11 | 26.10 | 10 | 19.12 | 4.12 | -1 |
| 3.10 | 18.9 | 4 | 11.11 | 27.10 | 10 | 20.12 | 5.12 | -1 |
| 4.10 | 19.9 | 5 | 12.11 | 28.10 | 10 | 21.12 | 6.12 | -1 |
| 5.10 | 20.9 | 5 | 13.11 | 29.10 | 10 | 22.12 | 7.12 | -2 |
| 6.10 | 21.9 | 5 | 14.11 | 30.10 | 10 | 23.12 | 8.12 | -2 |
| 7.10 | 22.9 | 5 | 15.11 | 31.10 | 10 | 24.12 | 9.12 | -3 |
| 8.10 | 23.9 | 5 | 16.11 | 1.11 | 9 | 25.12 | 10.12 | -3 |
| 9.10 | 24.9 | 6 | 17.11 | 2.11 | 9 | 26.12 | 11.12 | -4 |
| 10.10 | 25.9 | 6 | 18.11 | 3.11 | 9 | 27.12 | 12.12 | -4 |
| 11.10 | 26.9 | 6 | 19.11 | 4.11 | 9 | 28.12 | 13.12 | -5 |
| 12.10 | 27.9 | 6 | 20.11 | 5.11 | 9 | 29.12 | 14.12 | -5 |
| 13.10 | 28.9 | 7 | 21.11 | 6.11 | 9 | 30.12 | 15.12 | -6 |
| 14.10 | 29.9 | 7 | 22.11 | 7.11 | 9 | 31.12 | 16.12 | -6 |
| 15.10 | 30.9 | 7 | 23.11 | 8.11 | 8 | - | - | - |
| 16.10 | 1.10 | 7 | 24.11 | 9.11 | 8 | - | - | - |
| 17.10 | 2.10 | 7 | 25.11 | 10.11 | 8 | - | - | - |
| 18.10 | 3.10 | 8 | 26.11 | 11.11 | 8 | - | - | - |
| 19.10 | 4.10 | 8 | 27.11 | 12.11 | 8 | - | - | - |
| 20.10 | 5.10 | 8 | 28.11 | 13.11 | 7 | - | - | - |
| 21.10 | 6.10 | 8 | 29.11 | 14.11 | 7 | - | - | - |
| 22.10 | 7.10 | 8 | 30.11 | 15.11 | 7 | - | - | - |
| 23.10 | 8.10 | 9 | 1.12 | 16.11 | 6 | - | - | - |
| 24.10 | 9.10 | 9 | 2.12 | 17.11 | 6 | - | - | - |